\documentclass[epsfig,12pt]{article}
\usepackage{epsfig}
\usepackage{amsmath}
\usepackage{amsfonts}
\usepackage{graphicx}
\usepackage{cancel}
\usepackage{fullpage}
\usepackage[hidelinks]{hyperref}
\usepackage{appendix}

\newcommand{\Lagr}{\mathcal{L}}

\newcommand{\derOrd}[2]{\frac{d#1}{d#2}}
\newcommand{\der}[2]{\frac{\partial#1}{\partial#2}}

\newcommand{\non}{\nonumber\\}
\title{Sigma Model on a Squashed Sphere with a Wess-Zumino Term}
\date{}
\begin{document}

	\begin{titlepage}

		\begin{flushright}
			FTPI-MINN-20/01, UMN-TH-3911/20
			
		\end{flushright}
		
		\vspace{5mm}
		
		\begin{center}
			{  \Large \bf  
						Sigma model on a squashed sphere with a Wess-Zumino term
			}
			
			\vspace{7mm}

			{\large \bf   Daniel Schubring$^{\,a}$ and Mikhail~Shifman$^{\,b}$  }
			\end {center}
			
			\begin{center}
				$^a${\it  Physics Department,
					University of Minnesota,
					Minneapolis, MN 55455}\\
				
				$^b${\it  William I. Fine Theoretical Physics Institute,
					University of Minnesota,
					Minneapolis, MN 55455}
				
			\end{center}
			
			\vspace{10mm}
			
			\begin{center}
				{\large\bf Abstract}
			\end{center}
				A class of two-dimensional sigma models interpolating between $CP^1$ and the $SU(2)$ principal chiral model is discussed. We add the Wess-Zumino-Novikov-Witten (WZNW) term and examine the renormalization group flow of the two coupling constants which characterize the model under consideration. The model flows to the $SU(2)$ WZNW conformal field theory in the infrared (IR) limit. There is an ordinary phase in which the model flows from the {\em asymptotically free} ultraviolet (UV) limit of the $CP^1$ model coupled to an extra massless degree of freedom. At higher-loop order we find evidence that there is also a phase in which the model can flow from non-trivial fixed points in the UV. A non-perturbative confirmation of these extra fixed points would be desirable.

		\end{titlepage}
		
	\section{Introduction}
	\label{intro}
	
Nonlinear sigma models are often used as toy models in high energy physics to illuminate aspects of more realistic theories like QCD (see e.g. \cite{SigmaModelsQCD1984}). But even seemingly abstract models can turn out to be useful effective field theory descriptions of phenomena in both high energy physics and condensed matter. Some examples of the latter which directly pertain to the target space under consideration here are found in the study of frustrated spin systems \cite{frustratedReview}.
	
	The sigma model we are considering here is a deformation of both the $O(2N)$ model and the $CP^{N-1}$ model. The Lagrangian is given by
	\begin{align}
	\Lagr_\kappa=\frac{1}{2\lambda^2}\left(\partial n^\dagger \partial n -\kappa|n^\dagger\partial n|^2\right),\label{lagrNField}
	\end{align}
	where $n$ is an $N$-dimensional complex unit vector,
	$|n|^2=1$. When $\kappa=0$, this is just the $O(2N)$ model, i.e. the sigma model on $S^{2N-1}$. When $\kappa=1$, the model becomes $U(1)$ gauge invariant and is the sigma model on $CP^{N-1}$. For intermediate values of $\kappa$, the target space is a ``squashed sphere" which is topologically equivalent to $S^{2N-1}$, but which has a deformed metric along the $U(1)$ fibers of the fiber bundle defined by the natural map from $S^{2N-1}\rightarrow CP^{N-1}$.
	
	The $\kappa$ deformation breaks the global $O(2N)$ symmetry of $S^{2N-1}$ down to $SU(N)\times U(1)$, and so this sigma model is a natural candidate for an effective field theory of a system with such global symmetry. For this reason, it has been well studied in the condensed matter community beginning with the $N=2$ case in a 1989 paper by Dombre and Read on quantum antiferromagnets on a two-dimensional spatial triangular lattice \cite{DombreRead}. A 1995 paper by Azaria, Lecheminant, and Mouhanna study this model in both the weak coupling and large $N$ limits and contains many more references to work on this model from this era \cite{MouhannaEtAl1995}. A more recent 2018 paper considers this model as an effective field theory for antiferromagnets on a two-dimensional triangular lattice with noncoplanar ordering, and also on three-dimensional pyrochlore lattices \cite{BatistaEtAl2018}.
	
	From the high-energy point of view, this is also an interesting toy model, especially in two spacetime dimensions, which is connected to the high temperature limit of the condensed matter models in three spacetime dimensions. So far the introduction of an interpolating parameter $\kappa$ in the Lagrangian seems 
	rather {\em ad hoc}, but it arises naturally by coupling the $CP^{N-1}$ model to massless fields.
	
	To see this, note that the ordinary $CP^{N-1}$ Lagrangian can be written with an auxiliary gauge field $A_\mu$ in order to make the gauge symmetry obvious,
	\begin{align*}
	\frac{1}{2\lambda^2}\big(\partial_\mu +iA_\mu\big) n^\dagger\, \big(\partial^\mu -iA^\mu\big)n\,.
	\end{align*}
	The auxiliary gauge field can then be coupled to a massless Dirac fermion $\psi$,
	\begin{align}
	\frac{1}{2\lambda^2}\big(\partial_\mu +iA_\mu\big) n^\dagger\, \big(\partial^\mu -iA^\mu\big)n+\frac{1}{2\alpha}\bar{\psi}\gamma^\mu(i\partial_\mu +A_\mu)\psi,\label{lagrDirac}
	\end{align}
	where $\alpha$ is an arbitrary parameter we will eventually connect to $\kappa$. This Lagrangian in turn is connected to a Stueckelberg field $\phi$, through the bosonization map in two dimensions (2D),
	\begin{align}
	\frac{1}{2\lambda^2}\big(\partial_\mu +iA_\mu\big) n^\dagger\, \big(\partial^\mu -iA^\mu\big)n+\frac{1}{4\pi \alpha} (\partial_\mu \phi -A_\mu)^2. \label{lagrStueckelberg}
	\end{align}
	For an example of bosonizing a Dirac fermion in a background gauge field see e.g. \cite{FujikawaSuzuki2004}. The gauge can now be fixed so that $\phi=0$, and the additional term becomes a mass term for $A$ which breaks the confinement of the $CP^{N-1}$ model. When auxiliary $A$ field is integrated out we recover the squashed sphere Lagrangian \eqref{lagrNField} if we make the identification,
	\begin{align}
\alpha = \frac{1}{2\pi}\frac{\kappa\lambda^2}{1-\kappa}.\label{alpha}
	\end{align}
	So this formulation shows that the squashed sphere Lagrangian is something rather natural to consider, just being equivalent to a Higgsed $CP^{N-1}$. But going back to the original formulation \eqref{lagrNField} in terms of the parameter $\kappa$ emphasizes that it is somehow interpolating between the $CP^{N-1}$ and $O(2N)$ models, which is interesting because the properties of these models are very different. In particular $CP^{N-1}$ has a nontrivial second homotopy group and can be modified by a $\theta$ term, whereas this can not be defined in the case of the $O(2N)$ models. On the other hand, for $N=2$, the two components of the complex unit vector $n$ can be packaged into a matrix $U\in SU(2)$ where
	\begin{align*}
	U=\left(\begin{array}{cc}
		n_1^*& n_0\\
		-n_0^* & n_1
	\end{array}\right).
	\end{align*}
	In terms of $U$, the Lagrangian \eqref{lagrNField} becomes,
	\begin{align}
	\Lagr=\frac{1}{4\lambda^2}\text{Tr}\left(\partial^{\mu} U^{\dagger}\partial_\mu U\right)-\frac{\kappa}{2\lambda^2}\eta^{\mu\nu}I^{3}_\mu I^3_\nu,\label{lagrPCMkappa}
	\end{align}
	where
	$$I^3_\mu \equiv \frac{1}{2}\text{Tr}\left(-i U^\dagger \partial_\mu U \tau_3 \right),$$
	with $\tau_3$ the third Pauli matrix. Moreover, $\eta^{\mu\nu}$ is the spacetime (or worldsheet, depending on the interpretation) metric. In 2D Minkowski space
	$\eta^{\mu\nu}= {\rm diag}\{1,-1\}$ while in Euclidean $\eta^{\mu\nu}= {\rm diag}\{1,\,1\}$.
	
	The key point is that the first term is just the Lagrangian of $SU(2)$ principal chiral model (PCM), which makes sense since the target space $S^3$ of the $O(4)$ model is equivalent to the Lie group $SU(2)$.
	
	The principal chiral model can also be modified by a topological term: the Wess-Zumino-Novikov-Witten (WZNW) term \cite{WessZumino1971,Novikov1981,WittenWZW1984}. In this paper we will investigate the impact of the WZNW term on the squashed sphere sigma model, which is in some sense interpolating to the $CP^1$ model in which the WZNW term does not make sense in two spacetime dimensions.
	
	This is not the first time the $SU(2)$ PCM and the $CP^1$ model have been connected through topological terms. In the so-called Haldane conjecture \cite{HaldaneConjecture}, an antiferromagnetic Heisenberg spin chain with half-integer spin is shown to be equivalent to the $CP^1$ model with a theta term set to $\theta=\pi$. The antiferromagnetic Heisenberg spin chain in turn flows to a massless Dirac fermion in the IR, which can be equivalently represented as $SU(2)$ PCM with a WZNW term at level $k=1$. So the $CP^1$ model at $\theta=\pi$ flows to the $SU(2)$ WZNW model at level $k=1$ \cite{AffleckHaldane1987,ShankarRead1990}.
	
	However the squashed sphere model considered in the following is not the same as this flow. In the UV the model looks like $CP^1$ coupled to a massless fermion, as in \eqref{lagrDirac}. As usual this massless fermion will wash out any dependence on a theta term we might try to define. Even so, we can still define a WZNW term for the squashed sphere model or equivalently the Higgsed $CP^1$ model, and it will have some interesting consequences.

	Various generalizations of this squashed sphere model, both with and without a WZNW term, have been considered by a number of authors, particularly in regards to its integrability and applications to AdS/CFT (see e.g. \cite{TseytlinEtAl2014}). The classical integrability of the squashed sphere model was first shown by Cherednik in 1981 \cite{Cherednik1981}. A later rediscovery of the integrability \cite{KawaguchiYoshida2010} involves modifying the $SU(2)$ current by a topological current so as to preserve the flatness condition \cite{BIZJZ1979}. This result was also extended to the squashed sphere with a WZNW term, as we consider here, and the RG flow was calculated to one loop \cite{KawaguchiOrlandoYoshida2011,KawaguchiYoshida2014}.
	
	The one loop RG flow was also calculated for a four-parameter generalization of the squashed sphere target space \cite{Lukyanov2012} which in another limit reduces to the integrable sausage model \cite{SausageModel}. This four-parameter model and related target spaces were further considered in \cite{TseytlinEtAl2014}.
	
	Besides this four-parameter model, another generalization of the squashed sphere model is the Yang-Baxter model \cite{Klimcek2002} which is also integrable \cite{Klimcek2008}. The Yang-Baxter model with a WZNW term was considered \cite{DriezenEtAl2018}, and the classical flatness condition for the currents was shown to persist to one loop. Two-loop renormalizability of such models were recently addressed in \cite{dop1,dop2}.

	In this paper our focus is on the two-loop RG flow, which leads to new features not seen in the one loop case. With the second loop included,
	the RG flow drastically changes, in particular due to the emergence a second separatrix and one or more new nontrivial fixed points in the UV.
	In general, a tentative existence of an extra fixed point from
balancing one- and two-loop terms (with not necessarily small higher-order corrections) cannot not proven. As a counterexample we could refer to symmetric space models with negative curvature (but {\em without} the Wess-Zumino-Novikov-Witten term) \cite{Gubser}. Higher-loop corrections are small if the parameter $k\gg1$.
Since the phenomenon we detected\,\footnote{The additional fixed points show up at  $2<k\leq 8$, see Sec. 3.4.}  disappears at $k\geq 9$ we  can only hope that at $k$ smaller than 9, but not too small, (say, $k=7$, Fig. 2), the extra fixed points survive. To complete the proof non-perturbative methods are needed. 
	
	The paper is organized as follows. In Section \ref{section2} we will review briefly the squashed sphere model in the formulation \eqref{lagrPCMkappa} close to the $SU(2)$ PCM, and we will explain how to add a WZNW term. The discussion closely follows that of \cite{LeutwylerShifman1992}. The general results on renormalizing a sigma model with a WZNW term are reviewed briefly in Section \ref{sectionRenormalization}.
	
	The RG equations of the squashed sphere sigma model with a WZNW term are found and discussed in Section \ref{section3}. Section \ref{sectionBelow1stSeparatrix} discusses the ordinary regime in which the model flows to the $CP^1$ sigma model coupled to a massless fermion in the UV. Section \ref{sectionTesting} begins the discussion of the two loop results by comparing the loop expansion of the beta functions to the expansion about the WZNW CFT. In Section \ref{sectionExactTrajectory} a particular RG trajectory is found which is argued to be valid to all orders in perturbation theory. And finally in Section \ref{sectionNewFixedPoints}, non-trivial UV fixed points of the model are found at two loop order. The possible non-perturbative existence of these fixed points is further discussed in the conclusion, Section \ref{sectionDiscussion}.

	\section{Introducing the WZNW term}
	\label{section2}
	\subsection{Introducing the model}
	We will begin by recalling the Lagrangian of the $SU(N)$ principal chiral model (PCM),
	\begin{align}
	\Lagr_{\text{PCM}}=\frac{1}{4\lambda^2}\text{Tr}\left(\partial^{\mu} U^{\dagger}\partial_\mu U\right).\label{lagrPCM}
	\end{align}
	where $U$ is a $N\times N$ matrix in $SU(N)$. This has $SU(N)_L\times SU(N)_R$ global symmetry,
	 \begin{align*}
	 U(x)\rightarrow V_L U(x) V_R^{-1}.
	 \end{align*}
	 Up to a normalization factor, the Noether currents corresponding to these symmetries are,
	 \begin{align}
	 J_{L,\mu}= i \partial_\mu U U^\dagger,\qquad J_{R,\mu}=-i U^\dagger \partial_\mu U.\label{noetherPCM}
	 \end{align}
	 Of course, once additional terms are added to the action, the Noether currents $J_L, J_R$ will need to be modified, but these original combinations of $U$ matrices will still be useful. The Noether current under right isospin transformations here will also be referred to as $I$, and it may be expanded in terms of the Lie algebra basis $\tau_a$,
	 $$I_\mu = I^a_\mu\tau_a \equiv -i U^\dagger\partial_\mu U.$$
	 The Lie algebra basis is normalized as
	 $$\text{Tr}\left(\tau_a\tau_b\right)=2\delta_{ab}.$$
	 It will be convenient to reexpress the Lagrangian in terms of these currents.
	 	\begin{align}
	 	\Lagr_{PCM}=\frac{1}{2\lambda^2}\sum_a I^{a,\mu}I^a_{\mu}.
	 	\end{align}
	 Of course we could have equally well expanded in terms of $J_L$ instead of $J_R=I$, but as in \cite{SchubringShifman2019} we will explicitly break the $SU(N)_L\times SU(N)_R$ symmetry down to $SU(N)_L\times U(1)_R$ by adding extra terms depending on $J_R$.
	 
	 As far as eventually adding a WZNW term is concerned, the interesting case is when $N=2$, in which case the squashed sphere sigma model Lagrangian is just \cite{SchubringShifman2019}
	 \begin{align}
	 \Lagr=\frac{1}{2\lambda^2}\left((I^1)^2+(I^2)^2+(1-\kappa)(I^3)^2\right)\label{lagrSquashedSphere}
	 \end{align}
	 where the upper index taking values $1,2,3$ refers to the usual Pauli matrix basis of the Lie algebra. When $\kappa=0$ this is the ordinary $SU(2)$ PCM, which is equivalent to the sigma model on the sphere $S^3$. When $\kappa=1$, the $U(1)$ global symmetry becomes a gauge symmetry, and the model is equivalent to the sigma model on $CP^1$.
	 \subsection{Adding a WZNW term}
	 In two space-time dimensions, the only finite action field configurations are those which have a unique limit as the space-time argument goes to infinity, and thus these field configurations map out a two-dimensional surface in the squashed sphere target space homeomorphic to $S^3$. Since the target space is three-dimensional it is meaningful to consider the three-dimensional volume enclosed by the field configuration in the target space. A term proportional to this volume is exactly the WZNW term we will add to the action.
	 
	 This can be calculated by integrating over the volume form on the squashed sphere, thus we will need an expression for the determinant of the metric. In terms of an orthonormal basis on the target space expressed in terms of the vielbeins $e^a_\mu$, it is easy to show that the determinant of the metric is,
	 \begin{align}
	 \sqrt{g}=\epsilon^{\lambda\mu\nu}e^1_\lambda e^2_\mu e^3_\nu.
	 \end{align}
	 
	 The coordinates on the target space are not yet fixed, but they can be chosen to be compatible with the spacetime coordinates. The two-dimensional spacetime coordinates can be thought of as defining a coordinate system on the image of the field configuration $U(x)$ in the target space. And if we introduce an arbitrary third coordinate, the image of $U(x)$ can be continued throughout the bulk of the target space. This choice of coordinates is useful because the quantity $I^a_\mu$ can be interpreted as the projection of the coordinate vector $\partial_\mu$ onto the left-invariant vector field corresponding to the Lie algebra element $\tau_a$. And since the components of the target space metric can be read off the sigma model Lagrangian, our Lagrangian \eqref{lagrSquashedSphere} is telling us that the left-invariant vector fields $\tau_a$ are orthogonal. In other words, up to a constant normalization, the currents $I^a_\mu$ can be identified with the vielbeins.
	 \begin{align}
	 \sqrt{g}&=\frac{\sqrt{1-\kappa}}{\lambda^3}\epsilon^{\lambda\mu\nu}I^1_\lambda I^2_\mu I^3_\nu\non
	 &=\frac{-i}{12}\frac{\sqrt{1-\kappa}}{\lambda^3}\epsilon^{\lambda\mu\nu}\text{Tr}\left(I_\lambda I_\mu I_\nu\right).
	 \end{align}
	Now we will integrate this over the interior of the field configuration, but since the target space is homeomorphic to $S^3$ there is some ambiguity in which side is considered the interior and which is the exterior. We could even allow for the `interior' of the field configuration to wrap around the manifold multiple times. In any case, as usual the ambiguity in signed volume will be some integer multiple of the total volume of the target space, $2\pi^2 \lambda^{-3}\sqrt{1-\kappa}$. This ambiguity will be harmless in the path integral if we normalize it to be some integer multiple $k$ of $2\pi i$. So the WZNW action is just
	\begin{align}
	S_{\text{WZNW}}&=\frac{2\pi i k}{2\pi^2 \lambda^{-3}\sqrt{1-\kappa}}\int dx^3 \sqrt{g}\non
	&=\frac{k}{12\pi}\int dx^3\epsilon^{\lambda\mu\nu}\text{Tr}\left(I_\lambda I_\mu I_\nu\right).\label{WZWterm1}
	\end{align}
	Note that all dependence on the parameters $\lambda$ and $\kappa$ has canceled, so the WZNW term for the squashed sphere is exactly the same as for the unit sphere $S^3$.
	
	Adding the WZNW action to the original action of the PCM \eqref{lagrPCM}, the Noether currents are modified. Using the same normalization as in \eqref{noetherPCM}, they become,
	\begin{align}
	J_{L,\mu}= i \partial_\mu U U^\dagger-k\frac{\lambda^2}{2\pi}\epsilon_{\mu\nu}\partial_\nu U U^\dagger,\qquad J_{R,\mu}=-i U^\dagger \partial_\mu U-k\frac{\lambda^2}{2\pi}\epsilon_{\mu\nu}U^\dagger \partial_\nu U.\label{noetherWZW}
	\end{align}
	At a special value of the coupling constant which we will call $\lambda_k^2$,
	\begin{align}
	\lambda_k^2=\frac{2\pi}{k},
	\end{align}
	these currents reduce to one independent component in holomorphic coordinates, $z=x^0+ix^1$,
	\begin{align}
	J_{L,\bar{z}}= 2i \partial_{\bar{z}} U U^\dagger,\qquad J_{R,z}=-2i U^\dagger \partial_z U,\label{noetherWZWz}
	\end{align}
	and the Noether current conservation law implies that $J_{R,z}$ only depends on $z$, and $J_{L,\bar{z}}$ only depends on $\bar{z}$. These currents form a Kac-Moody algebra of level $k$ and this is of course just the WZNW conformal fixed point first found in \cite{WittenWZW1984}.
	
	Since we will be modifying the Lagrangian by terms of the form $I^\mu I_\mu \propto I_z I_{\bar{z}}$ we will need to consider not only $I_z$, which is one component of a conserved current, but also $I_{\bar{z}}$ which is no longer conserved and thus can have an anomalous dimension. The scaling dimension of this operator was calculated by Knizhnik and Zamolodchikov \cite{KnizhnikZamolodchikov1984}. It takes the form $2\Delta_1+1$, where in case of the group $SU(2)$,
	\begin{align}
	\Delta_1 = \frac{2}{2+k}.\label{Idimension}
	\end{align}
	
	\subsection{Renormalizing WZNW models}\label{sectionRenormalization}
	
	The problem of renormalizing sigma models with a WZNW term has been considered by a number of authors \cite{BraatenEtAl1985}\cite{BosWZW1987}\cite{Ketov2Loop1987}. The key step is to rewrite the WZNW term in the action as a two-dimensional integral on the same footing as the ordinary sigma action and then apply the same background field method which works to find the renormalization of the sigma model without a WZNW term (an early example of which is found in \cite{HonerkampEcker}). Note that although the two-loop beta function would be expected to depend on a choice of renormalization scheme already at two-loops \cite{MT}, in Sec.~\ref{sectionTesting} we will test that this scheme is valid by matching to conformal perturbation theory at the WZW fixed point.
	
	The WZNW term in the action \eqref{WZWterm1} may be rewritten as
	\begin{align}
S_{\text{WZNW}}&=\frac{i}{3}\int dx^3 S_{abc}\epsilon^{\lambda\mu\nu}\partial_\lambda\phi^a\partial_\mu\phi^b\partial_\nu\phi^c\label{WZWterm2}
	\end{align}
	where $\phi^a$ are the fields in the sigma model mapping to coordinates in the target space, and $S_{abc}$ is proportional to the volume form on the target space $\omega_{abc}$, \begin{align}
	S_{abc}=\frac{k\lambda^3}{2\pi\sqrt{1-k}}\omega_{abc}.
	\end{align}
	Of course since $S$ is a three-form on a three-dimensional manifold, it is a closed form. And even on higher dimensional target spaces, the WZNW term may be written as an integral over a closed form. This means that locally (but not globally), we may write $S$ as the exterior derivative of a 2-form gauge field $h$,
	$$S_{abc}=\partial_{[a}h_{bc]},$$
	and apply Stokes' theorem to rewrite the action as
	\begin{align}
S_{\text{WZNW}}&=\frac{i}{3}\int dx^2 h_{ab}\epsilon^{\mu\nu}\partial_\mu\phi^a\partial_\nu\phi^b.
	\end{align}
	which just looks like an antisymmetric version of the ordinary sigma model action,
	\begin{align}
S_{{\sigma}}&=-\frac{1}{2}\int dx^2 g_{ab}\eta^{\mu\nu}\partial_\mu\phi^a\partial_\nu\phi^b,
	\end{align}
	where $g$ is the target space metric and $\eta^{\mu\nu}$ is defined after Eq. (\ref{lagrPCMkappa}). Note that in our conventions all $\lambda$ and $\kappa$ dependence is absorbed into the definitions of $g$ and $h$.
	
	The problem of finding the renormalization of a sigma model with general $g$ and $h$ has been solved up to three loops \cite{KetovEtAl3Loop1990}, although there are still some ambiguities left to be cleared up, as we will discuss later. In our case, $S$ is proportional to the volume form, so its covariant derivative vanishes, and it also satisfies the identity
	$$S_{[ab}^{\quad h}S_{c]gh}=0.$$
	This will simplify some of the formulas for the $\beta$ functions slightly. To two loops the beta function is given\,\footnote{The expression for 
	the derivative of the $\beta$ function in  (\ref{betaFunctionGeneral})
generally speaking must involve  symmetrization in $a,b$ on the right-hand side. However, in the case under consideration
in which  the
antisymmetric tensor $h_{ab}$  does not run  (corresponding to the WZNW
term) this is irrelevant,  as discussed above Eq. (\ref{betaFunctionGeneral}).} as \cite{Ketov2Loop1987}
	\begin{align}
	\mu\der{}{\mu}g_{ab}=\frac{1}{2\pi}\hat{R}^{c}_{\,\,abc}+\frac{1}{8\pi^2}\hat{R}_{acdf}\hat{R}_{b}^{\,\,cdf}-\frac{1}{(2\pi)^2}\hat{R}_{adfb}S^{d}_{\,\,gh}S^{fgh},\label{betaFunctionGeneral}
	\end{align}
	where $\hat{R}_{abcd}$ is the Riemann curvature tensor $R$ modified by $S$ in such a way that it has an interpretation as a Riemann curvature for a target space metric with torsion \cite{BraatenEtAl1985},
	\begin{align}
	\hat{R}_{abcd}\equiv R_{abcd}-S_{fab}S^f_{\,\,dc}.
	\end{align}
	The last term in \eqref{betaFunctionGeneral} involved some ambiguities in continuing the Levi-Civita tensor in dimensional regularization which were fixed by matching the beta function to the dimension of the operator perturbing the conformal fixed point \cite{BosWZW1987}.

	For the ordinary ($\kappa=0$) $SU(2)$ WZNW model,
	$$R_{abcd}=\lambda^2\left(g_{ad}g_{bc}-g_{ac}g_{bd}\right),$$
	and given that $S$ is proportional to the Levi-Civita tensor
	the contraction $S_{fab}S^f_{\,\,dc}$ produces an identical structure. So the full Riemann curvature with torsion $\hat{R}$ is
	$$\hat{R}_{abcd}=\lambda^2(1-\eta^2)\left(g_{ad}g_{bc}-g_{ac}g_{bd}\right),$$
	where we have defined the useful parameter\,\footnote{The parameter $\eta$ in (\ref{peta}) and below (without indices) is not to be confused
	with the spacetime metric $\eta^{\mu\nu}$.}
	\begin{align}
	\eta \equiv \frac{k\lambda^2}{2\pi}=\frac{\lambda^2}{\lambda_k^2}.
	\label{peta}
	\end{align}
	When we are at the point $\eta=1$, the modified Riemann curvature $\hat{R}_{abcd}=0$ and thus the $\beta$ function vanishes. And this is of course just the point where $\lambda^2=\lambda_k^2$ which was introduced above in the context of the current algebra.
	
	\section{RG flow of the squashed sphere with WZNW term}\label{section3}
	
	It is now a simple matter to calculate the $\beta$ function when $\kappa\neq 0$. Now the Riemann tensor will no longer take a form proportional to $\left(g_{ad}g_{bc}-g_{ac}g_{bd}\right)$, but it can still be calculated from the structure coefficients of the group as in e.g. \cite{MouhannaEtAl1995,SchubringShifman2019}. Taking the left invariant vector fields corresponding to the standard Pauli matrix basis for the Lie algebra as a basis for the tangent space, the metric is
	\begin{align}
	g_{11}=g_{22}=\frac{1}{\lambda^2},\qquad g_{33}=\frac{1-\kappa}{\lambda^2}
	\end{align}
	and the Riemann tensor $R_{abcd}=R_{[ab][cd]}$ is
	\begin{align}
	R_{1212}=-\frac{1}{\lambda^2}(1+3\kappa),\qquad R_{1313}=R_{2323}=-\frac{1}{\lambda^2}(1-\kappa)^2.
	\end{align}
	Also recall that the antisymmetric $S$ tensor is proportional to the volume form $\omega$,
	\begin{align}
	S_{abc}=\frac{\lambda \eta}{\sqrt{1-k}}\omega_{abc}.
	\end{align}
Applying the general two-loop RG equations \eqref{betaFunctionGeneral}, we find beta functions for $\lambda$ and $\kappa$,
	\begin{align}\!\!
	\mu \derOrd{}{\mu}\left(\frac{1}{\lambda^2}\right)&=\frac{1}{\pi}\left[(1+\kappa)-\frac{1}{1-\kappa}\eta^2\right]+\frac{\eta}{\pi k}\left[1+2\kappa+5\kappa^2-4\frac{1+\kappa}{1-\kappa}\eta^2+3\frac{1}{(1-\kappa)^2}\eta^4\right],\non\non
	\mu \derOrd{}{\mu}\left(\frac{1-\kappa}{\lambda^2}\right)&=\frac{1}{\pi}\left[(1-\kappa)^2-\eta^2\right]+\frac{\eta}{\pi k}\left[(1-\kappa)^3-4(1-\kappa)\eta^2+3\frac{1}{(1-\kappa)}\eta^4\right].
	\label{RGequations}
	\end{align}
	When $\kappa=0$ this agrees with the usual two-loop RG equation for the WZNW model \cite{BosWZW1987,Ketov2Loop1987}. When $\kappa\neq 0$ but $k=0$, which implies $\eta=0$ and $\eta/k= \lambda^2/2\pi$, then the RG equations agree with the two-loop equations first found in \cite{MouhannaEtAl1995}.

	\begin{figure}[h]
	\begin{center}
		\includegraphics[width=7cm]{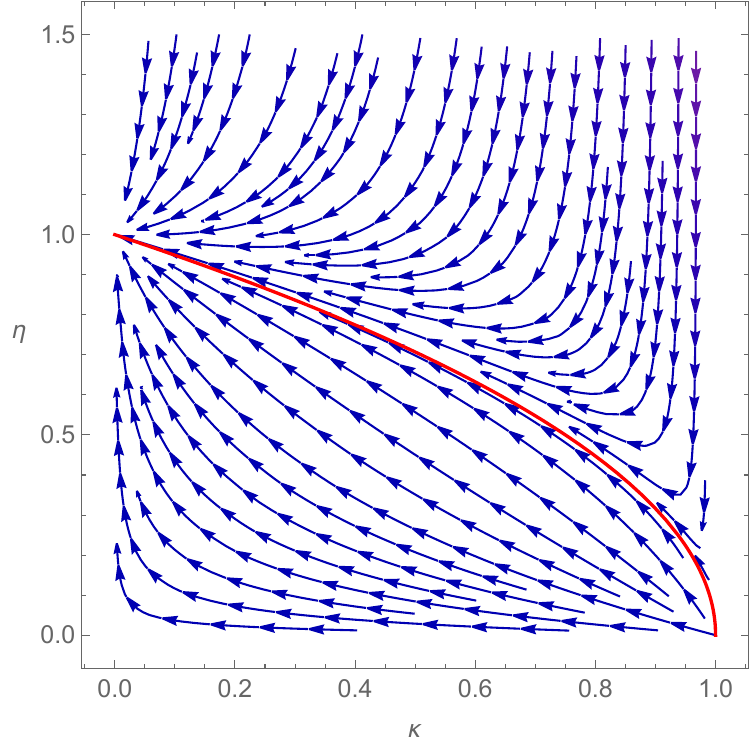}
		\end{center}
		\caption{\small RG flow for the squashed sphere sigma model with a WZNW term to one loop. The flow is pointing towards the IR. The separatrix $\eta=\sqrt{1-\kappa}$ is plotted in red.}\label{Fig1loop}
	\end{figure}
	
	\subsection{Below the first separatrix}\label{sectionBelow1stSeparatrix}
	
	Let us first consider the 1-loop behavior. The RG flow is plotted in Fig. \ref{Fig1loop}. Everything flows to the WZNW CFT in the IR. The curve $\eta=\sqrt{1-\kappa}$ solves the RG equations, and is in fact a separatrix determining two classes of UV behavior. Below the separatrix every trajectory flows in the UV to the asymptotically free fixed point related to $CP^1$. And above the separatrix the coupling constant $\eta$ and thus $\lambda^2$ appears to diverge in the UV. This divergence is quite possibly an artifact of taking only a finite order in perturbation theory, but we will have more to say on this later.

	Actually, as shown in \cite{KawaguchiOrlandoYoshida2011} and \cite{DriezenEtAl2018}, the separatrix $\eta=\sqrt{1-\kappa}$ is exactly the condition needed for the $SU(2)$ currents at non-zero $\kappa$ to satisfy the flatness condition of \cite{BIZJZ1979} without modification by a topological current. The condition $\eta=\sqrt{1-\kappa}$ solving the RG equations to one loop can be seen as a specific case of the preservation of the classical integrability condition for Yang-Baxter models with a WZNW term to one loop as found in \cite{DriezenEtAl2018}. Note that the exact condition $\eta=\sqrt{1-\kappa}$ is no longer preserved by the RG equations to two loops, although there is still a separatrix which approximately satisfies this for large $k$. 
	
	Far below the separatrix, for $\eta\ll\sqrt{1-\kappa}$, the RG equations reduce to the RG equations for the squashed sphere sigma model without a WZNW term. There is an RG invariant in this case first found in \cite{MouhannaEtAl1995},
	$$K=\frac{\sqrt{\kappa}}{1-\kappa}\lambda^2.$$
	
	As pointed out in \cite{SchubringShifman2019}, this value $K$ is the exponent of power law behavior of two-point functions of the unit vector $n$ field in the UV. This makes sense considering the representation of the squashed sphere model in terms of a Stueckelberg field $\phi$ discussed earlier, see equation \eqref{lagrStueckelberg}. In the UV, where $\kappa\approx 1$ and $\lambda\approx 0$, the correlation function is being dominated by a phase factor $\exp(i\phi)$, where $\phi$ is essentially a massless free field. In terms of the parameter $\alpha$ in \eqref{alpha}, this has two-point function,
	$$\langle e^{-i\phi(z)}e^{i\phi(0)}\rangle\propto |z|^{-\alpha}\approx |z|^{-\frac{K}{2\pi}} $$
	which is exactly the value of the exponent found in \cite{SchubringShifman2019}.
	
	Note in passing that we had been considering $\phi$ in \eqref{lagrStueckelberg} as a Stueckelberg field coupled to the gauge field $A$, but we can fix the gauge by making one of the components of $n$ real, and then integrate out $A$. In doing so, the sigma model Lagrangian takes a form where $\phi$ has the interpretation as the coordinate along the $U(1)$ fibers of the squashed sphere target space in a coordinate system closely related to the Fubini-Study coordinates (detailed in the appendix of \cite{SchubringShifman2019}, see also \cite{SoftTheorem}). The coupling between $\phi$ and the $CP^1$ degrees of freedom involves an extra factor of $\lambda^2$ and so the $\phi$ field decouples in the $\kappa\rightarrow 1, \lambda^2\rightarrow 0$ limit.
	
	This means that the squashed sphere model at $\kappa\approx 1$ is not a small perturbation on $CP^{1}$ alone, the additional degree of freedom given by $\phi$ is very important to the behavior of the correlation functions. This can also be seen by considering the Zamolodchikov c-theorem \cite{CTheorem}. The central charge at the WZNW fixed point is
	$$c=\frac{3k}{k+2}$$
	and since the central charge monotonically decreases along the RG flow and we are free to take $k$ as large as we want, there must be a central charge of at least $3$ in the UV. This emphasizes the point that the theory near $\kappa=1$ can not be just the $CP^1$ sigma model, which has a central charge of $2$ at its asymptotically free UV fixed point. In order for the flow to be consistent we need the extra degree of freedom given by the $\phi$ field.
	
	This is in contrast to the situation with the Haldane conjecture \cite{HaldaneConjecture,AffleckHaldane1987}. There we are considering $CP^1({\theta=\pi})$ flowing to only the level $k=1$ WZNW model, which has central charge $c=1$. So, as pointed out in \cite{ShankarRead1990}, it is perfectly consistent for the UV behavior to be given by the asymptotically free fixed point of the $CP^1$ model alone.
	
	Returning now to our discussion of $K$ and the correlation functions, note that even with the WZNW term, the quantity $K$ has an unambiguous definition below the separatrix through the limit of $\lambda^2/(1-\kappa)$ in the UV. Correlation functions of $n$ now have power law behavior in both the UV and IR. As just mentioned, the UV behavior is determined by the value of $K$, and the IR behavior is given by the scaling dimension of $n$ field in the WZNW CFT. Since the field $n$ is just components of $U$, which is a primary field in the WZNW CFT, it has scaling dimension 2$\Delta_{1/2}$, where  $\Delta_{1/2}$ was calculated in \cite{KnizhnikZamolodchikov1984},
	$$\Delta_{1/2}=\frac{3/4}{k+2}.$$
	Furthermore, as in the squashed sphere model without a WZNW term, trajectories with small values of $K$ pass near the $\kappa=0,\,\, \eta=0$ fixed point and so display crossover behavior associated to the UV behavior of the SU(2) PCM.
	
	\subsection{Testing the loop expansion}\label{sectionTesting}
	
	The RG equations \eqref{RGequations} are based on a loop expansion of the action. Our model has three parameters, $\lambda^2$ (or equivalently $\eta$), $\kappa$, and the discrete parameter $k$. So it perhaps is not clear at first which small parameter we are expanding in. But recall that the action  \eqref{lagrSquashedSphere} implies
	\begin{align*}
	S=\frac{k}{4\pi\eta}\int dx^2\big[(I^1)^2+(I^2)^2+(1-\kappa)(I^3)^2\big]+\frac{k}{12\pi}\int dx^3\epsilon^{\lambda\mu\nu}\text{Tr}\left(I_\lambda I_\mu I_\nu\right).
	\end{align*}
	Since $k^{-1}$ multiplies each term in the same position as $\hbar$, it should be treated as the loop expansion parameter, and indeed each term of the RG equations \eqref{RGequations} has a definite power of $k^{-1}$. At large $k$ successive loop contributions are suppressed.
	
	As a consistency check this expansion in $k^{-1}$ can be compared with a perturbative expansion about the WZNW fixed point action $S_0$, in a calculation similar to the $\kappa=0$ case \cite{BosWZW1987,Ketov2Loop1987}. If the CFT action $S_0$ is perturbed by operators $\mathcal{O}^i$ each of which has a well-defined scaling dimension $\Delta_i$,
	$$S=S_0 + g_i \mathcal{O}^i,$$
	then under a RG coarse graining from $\mu$ to $\mu'$, to lowest order in the small parameter $g_i$ the action will transform to
	$$S_0+g_i \left(\frac{\mu'}{\mu}\right)^{\Delta_i} \mathcal{O}^i\approx S_0+g_i \left(1+\Delta_i \log\frac{\mu'}{\mu}\right) \mathcal{O}^i.$$
	So to the lowest order the $\beta$ function for $g_i$ is just,
	\begin{align}
	\beta_{g_i}=g_i \Delta_i + \mathcal{O}(g^2).\label{betaFirstOrder}
	\end{align}
	In our case the total action is,
	\begin{align*}
	S= S_0 + \frac{k}{2\pi}\left(\frac{1-\eta}{\eta}\right)\int dx^2\frac{1}{2}\left[(I^1)^2+(I^2)^2\right]+\frac{k}{2\pi}\left(\frac{1-\kappa-\eta}{\eta}\right)\int dx^2\frac{1}{2}(I^3)^2.
	\end{align*}
	So the coefficients of the operators perturbing the fixed point action are
	\begin{align}
	g_\eta\equiv\frac{k}{2\pi}\left(\frac{1-\eta}{\eta}\right),\qquad g_\kappa \equiv \frac{k}{2\pi}\left(\frac{1-\kappa-\eta}{\eta}\right).\label{gParameters}
	\end{align}
	Expanding the beta functions \eqref{RGequations} in these parameters,
	\begin{align}
	\mu \derOrd{}{\mu}\left(\frac{1}{\lambda^2}\right)=\left(\frac{4}{k}-\frac{8}{k^2}\right)g_\eta+\mathcal{O}(g^2),\qquad \derOrd{}{\mu}\left(\frac{1-\kappa}{\lambda^2}\right)=\left(\frac{4}{k}-\frac{8}{k^2}\right)g_\kappa+\mathcal{O}(g^2)\label{RGdimension}
	\end{align}
	And indeed, using the dimension $2\Delta_1+1$ of the operator $I_{\bar{z}}$ in \eqref{Idimension}, the dimension of the operators $\int dx^2(I^a)^2$ is
		$$2\Delta_1=\frac{4}{2+k}=\frac{4}{k}-\frac{8}{k^2}+\mathcal{O}(k^{-3}).$$
	So the loop expansion to two loops agrees with conformal perturbation theory to first order.
	\subsection{An exact RG trajectory}\label{sectionExactTrajectory}
	Notice that the straight line $\eta=1-\kappa$ satisfies the RG equations \eqref{RGequations} to two loops. The meaning of this may be clarified by considering the parameters perturbing the WZNW fixed point \eqref{gParameters}. If $g_\kappa$ is set to zero at one scale, it remains zero under the RG flow. No $\left(I^3\right)^2$ perturbation on the fixed point is generated through renormalization.\footnote{After our publication the interesting paper \cite{levine2021integrability} appeared which introduces the so-called $T^{1,q}$ model. The exact RG trajectory in this section appears as the intersection of two parameter constraints where the $T^{1,q}$ model is classically integrable.}
	
	In fact, we conjecture that this will hold not just to two loops, but to all orders in perturbation theory. The reason that $\eta=1-\kappa$ satisfies the RG equations is that the $\beta$ function for $(1-\kappa)/\lambda^2$ (or equivalently, for $g_\kappa$) is of the form \eqref{RGequations}
	\begin{eqnarray}
	\mu \dfrac{d}{d\mu}\left(\frac{1-\kappa}{\lambda^2}\right)&=&\frac{1}{k^0}\left[a^{0}_{2,0}(1-\kappa)^2+a^{0}_{0,2}\eta^2\right]
	+\frac{1}{ k^1}\left[a^{1}_{3,1}(1-\kappa)^3\right.
	\nonumber\\[2mm]
	&+& \left. a^{1}_{1,3}(1-\kappa)\eta^3+a^{1}_{-1,5}(1-\kappa)^{-1}\eta^5\right].\nonumber
	\end{eqnarray}
	Here $a^n_{j,k}$ is just notation for the numerical coefficient of the term $(1-\kappa)^j\eta^k$ in the $(n-1)$-loop correction to the beta function. Since $\kappa=0,\,\, \eta=1$ is a fixed point, we must have the condition
	\begin{align}
	{\sum_{i,j}a^n_{i,j}=0}.\label{condition1}
	\end{align}
	So, given this condition, the reason why $\eta=(1-\kappa)$ solves the RG equations is because at each loop order $n$, the sum of the powers of $\eta$ and $(1-\kappa)$ is the same for each term. In other words, for each $n$, $i+j$ is constant for all non-vanishing $a^n_{i,j}$.
	
	But consider now the general expansion of the beta function in the loop index $n$ and as a Legendre expansion in $(1-\kappa)$ and $\eta$,
	 $$\mu \derOrd{}{\mu}\left(\frac{1-\kappa}{\lambda^2}\right)=\sum_{n,i,j}\frac{1}{k^n}a^{n}_{i,j}(1-\kappa)^i\eta^j.$$
	 Expanding to first order in $g_\eta,g_\kappa$ we have,
	 $$\mu \derOrd{}{\mu}\left(\frac{1-\kappa}{\lambda^2}\right)=\sum_{n}\frac{2\pi}{k^{n+1}}\left(g_\kappa\sum_{i,j}ia_{i,j}^n +g_\eta\sum_{i,j}(i+j)a_{i,j}^n \right)+\mathcal{O}(g^2).$$
	 But as in \eqref{betaFirstOrder} there should be no first order $g_\eta$ term. So besides \eqref{condition1}, there is also the condition
	 \begin{align}
{\sum_{i,j}(i+j)a_{i,j}^n=0}.\label{condition2}
	 \end{align}
	 This condition can also be derived by considering the $1/\lambda^2$ beta function at $\kappa=0$ and demanding that the dimension agrees with the $(1-\kappa)/\lambda^2$ beta function at $g_\eta=0$, as in \eqref{RGdimension}.
	 
	 The point is now that a simple way to satisfy this second condition \eqref{condition2}, is if $i+j$ is a constant for all non-vanishing $a^n_{i,j}$, in which case it simply reduces to the first condition \eqref{condition1}. And as mentioned above, this is exactly what is needed for $\eta=1-\kappa$ to solve the RG equations. So while this is not a proof, it seems rather plausible that $\eta=1-\kappa$, or equivalently $g_\kappa=0$, is preserved under the RG flow to all orders.

	\subsection{Additional fixed points}\label{sectionNewFixedPoints}
	
	\begin{figure}[h]
	\begin{center}
		\includegraphics[width=7cm]{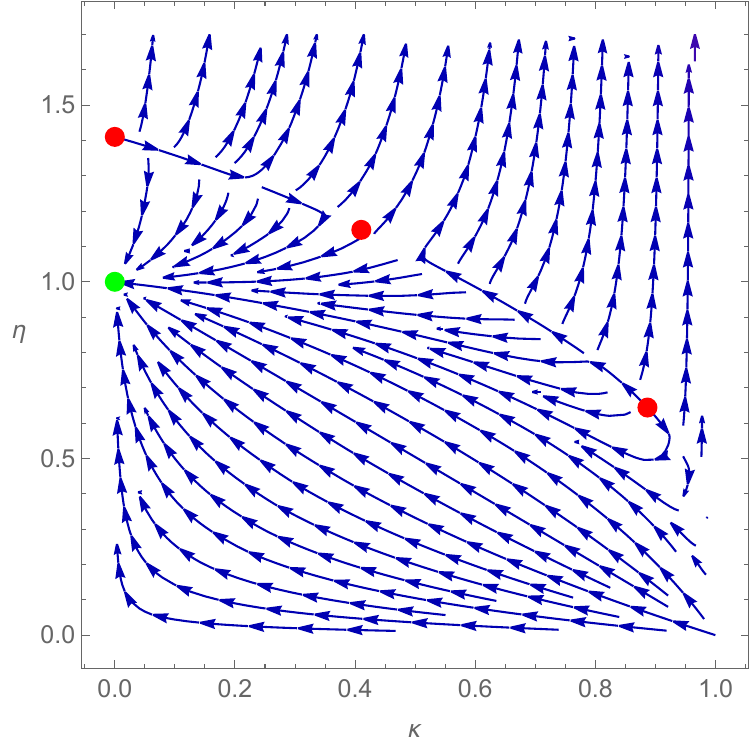}\qquad\includegraphics[width=7cm]{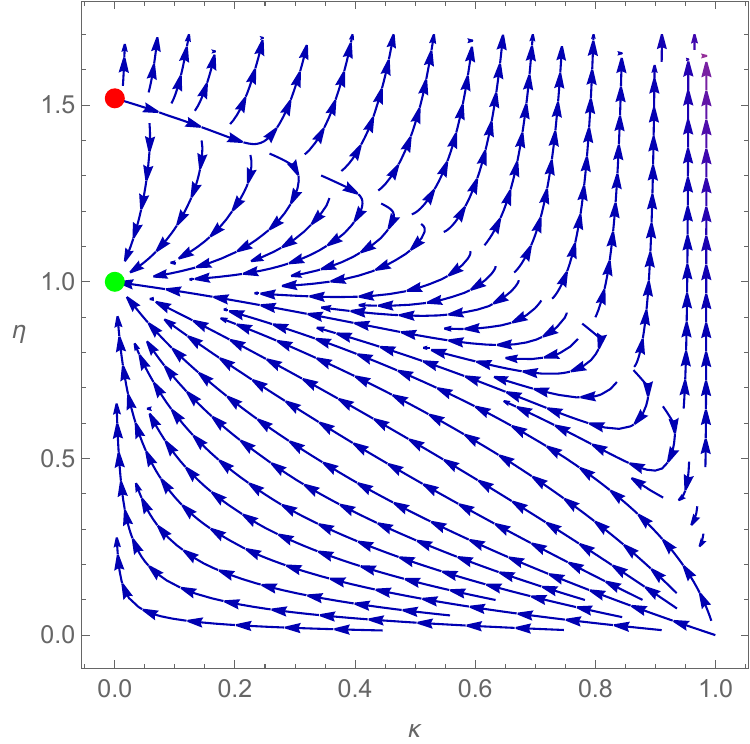}
		\end{center}
		\caption{\small RG flow for the squashed sphere sigma model with a WZNW term to two loops. On the left is the level $k=7$ and on the right is $k=9$. The green dot is the ordinary WZNW CFT, and the red dots are additional fixed points. The two fixed points with $\kappa>0$ disappear for levels $k\geq 9$.}\label{Fig2loop}
	\end{figure}

	In Fig. \ref{Fig2loop}, the RG flow is plotted to two loops. The behavior below the first separatrix, which is now a slight deformation of the one loop result $\eta=\sqrt{1-\kappa}$, is qualitatively the same as for one loop, but the behavior above the first separatrix is dramatically different. There is now a second separatrix, and between the first and second separatrices the trajectories flow to the WZNW CFT in the IR, but flow to one or more new nontrivial fixed points in the UV, plotted in red in Fig. \ref{Fig2loop}.
	
	Where the second separatrix meets $\kappa=0$ there is an IR unstable fixed point. Setting $\kappa=0$ in the RG equations \eqref{RGequations}, and dividing out the factor of $(1-\eta^2)$, we see that the unstable fixed point is at the value $\eta=\eta_0$ which is the real root of the cubic equation
	\begin{align}
	3\eta_0^3-\eta_0-k=0.\label{cubicFixedPoint}
	\end{align} 
	When $k$ is in the range $2<k\leq 8$, there are also two additional fixed points for $\kappa>0$, as shown on the left of Fig. \ref{Fig2loop}. The coordinates of these fixed points can be found by first noting that $\eta=1-\kappa$ solves the RG equations, as noted in Section \ref{sectionExactTrajectory}. This means $(1-\kappa)-\eta$ should be a factor in the RG equation \eqref{RGequations} for $((1-\kappa)/\lambda^2),$ and in fact, we can factor out $(1-\kappa)^2-\eta^2,$
	\begin{align*}
	\mu \derOrd{}{\mu}\left(\frac{1-\kappa}{\lambda^2}\right)&=\frac{(1-\kappa)^2-\eta^2}{\pi k(1-\kappa)}\left[k(1-\kappa)+\eta(1-\kappa)^2-3\eta^3\right].
	\end{align*}
	And we can also solve for the RG equation of $\kappa$ alone,
	\begin{align*}
	\mu \derOrd{\kappa}{\mu}&=-\frac{4\eta}{k}\kappa\left[(1-\kappa) +\frac{2\eta}{k}\left((1-\kappa^2) -2 \eta^2\right)\right].
	\end{align*}
	The fixed points $(\kappa_0,\eta_0)$ are roots of the bracketed polynomials on the RHS of these two equations, and these can be further manipulated so that $\kappa_0$ is given by the two real roots of a single quartic equation, and then $\eta_0$ can be easily found from the value of $\kappa_0$,
	\begin{align}
4(1-\kappa_0)&(1+3\kappa_0)(1+5\kappa_0)^2-k^2=0,\\
\eta_0&=\frac{k}{10\kappa_0+2}.
	\end{align}
	This quartic equation also indicates an upper bound on $k$ for which these two extra fixed points exist. At $k\geq 9$, these two fixed points disappear as in the right side of Fig. \ref{Fig2loop}, and since the loop expansion is essentially an expansion in $k^{-1}$, it is not clear whether these two fixed points are an artifact of the loop expansion or not.
	\section{Discussion of results}\label{sectionDiscussion}
	In fact it is not clear whether the new unstable fixed point at $\kappa=0$ which solves \eqref{cubicFixedPoint} is an artifact or not. Of course at $\kappa=0$ this is just the ordinary well-known SU(2) WZNW model, so it is perhaps surprising that there would be an additional fixed point. But this fixed point does already appear in the two-loop RG equations published long ago \cite{BosWZW1987,Ketov2Loop1987}.
	
	Note that the loop expansion may be expected to be more accurate at high $k$, and according to \eqref{cubicFixedPoint} the $\eta$ coordinate of this fixed point scales as $\eta_0\sim k^{1/3}$ at large $k$. Although $\eta$ increases with $k$, the parameter $\eta$ actually multiplies the action in the combination $2\pi\eta/k=\lambda^2$. And since $\lambda^2$ of the unstable fixed point scales as $k^{-2/3}$, naively it appears to be in the weak-coupling regime when $k$ is large.
	
	There are actually three-loop results available that can illuminate the issue \cite{KetovEtAl3Loop1990}. The $ \beta$ function at $\kappa=0$  is\,\footnote{As was mentioned in Sec. \ref{sectionRenormalization}, the RG  scheme dependence ambiguity at three loops has not yet been resolved
(for a related discussion in the
pure-metric case see, e.g., Appendix in \cite{dop1}).}
	\begin{eqnarray}
	\mu \frac{d}{d\mu}\left(\frac{1}{\lambda^2}\right)&=&\frac{1}{\pi}\left(1-\eta^2\right)+\frac{\eta}{\pi k}\left(1-\eta^2\right)\left(1-3\eta^2\right)
	\nonumber\\[2mm]
	&+&\frac{\eta^2}{2\pi k^2}(1-\eta^2)\Big[8+(1-\eta^2)q_2+(1-\eta^4)q_4\Big].
	\label{dop3}
	\end{eqnarray}
The	numerical parameters $q_2$ and $q_4$ in (\ref{dop3}) are scheme-dependent; in particular, they  depend on the dimensional regularization of the Levi-Civita symbol. The authors of \cite{KetovEtAl3Loop1990}  were unable to  fix them by first-order perturbation theory about the WZNW CFT.  So they are somewhat questionable and below we will use the values suggested in \cite{KetovEtAl3Loop1990} only for the purpose of orientation,
	\begin{align}
	q_2=-\frac{10}{3},\qquad q_4=-\frac{5}{3}.
	\end{align}
	With the above remark in mind, we will nevertheless  have a look on  whether the unstable fixed point \eqref{cubicFixedPoint} survives after including the three-loop terms. Dividing out the $1-\eta^2$ factor associated with the WZNW fixed point, and keeping only the highest order terms in $\eta$ at each loop, we have the fixed points as solutions to the equation
	$$-\frac{q_4}{2k}\eta^6-3\eta^3+k=0\,.$$
	The unstable fixed point survives for large $k$ as long as $q_4>-9/2$, which is satisfied if the value given in \cite{KetovEtAl3Loop1990} is correct. What is more, given $q_4<0$, there is a second fixed point which is stable in the IR and first appears at three loops.
	
	Although these three-loop results are only for $\kappa=0$, let us consider what this might imply about the $\kappa>0$ behavior. As seen in Fig. \ref{Fig2loop}, at the order of two loops, all RG trajectories above the second separatrix flow to strong coupling in the IR. Now given there is an IR stable fixed point that appears at three loops, it is very tempting to conjecture that the new stable fixed point at three loops lies at the end of a new third separatrix, above which all trajectories flow to strong coupling in the UV. If the pattern continues, at each loop order there could be a new separatrix and a new fixed point at $\kappa=0$, alternating between IR stable and unstable. 
	
	Needless to say, this is a shaky argument since not even the unstable fixed point at $\kappa=0$ appearing at two loops has been proven to exist outside of perturbation theory. Given the value of $q_4$ in \cite{KetovEtAl3Loop1990}, the three loop correction only shifts the fixed point value $\eta_0$  by a numerically small amount, but the correction is also of order $k^{1/3}$. To investigate this it might be useful to go beyond perturbation theory in $k^{-1}$, and compare to conformal perturbation theory about the WZNW CFT at higher orders in $g_\eta, g_\kappa$.
	
	\vspace{2mm}

	To summarize, in this paper we studied a two-dimensional field theory which is a deformation of the $SU(2)$ WZNW model that explicitly breaks one of the two $SU(2)$ global symmetries in the UV. The RG flow is partitioned by separatrices. Below the first separatrix, all trajectories flow in the UV to the weak-coupling fixed point of the $CP^1$ sigma model coupled to a massless fermion or Stueckelberg field. Above the first separatrix, to lowest order all trajectories appear to flow to a Landau pole in the UV, but as higher loops are added these trajectories appear to instead flow to new ``asymptotically safe" non-trivial fixed points. Demonstrating the non-perturbative existence of these fixed points, and the question of what CFT they correspond to is still an open question. 
	
	\section*{Acknowledgments}
	We would like to thank Nat Levine, Arkady Tseytlin, and 
	Sibylle Driezen for explaining the connection to their work and providing many helpful references.  We are also grateful to Sergey V. Ketov and Alexander A. Voronov for useful discussions. This work is supported in part by DOE grant DE-SC0011842.

\end{document}